\documentclass[sigconf]{acmart}

\usepackage[utf8]{inputenc} % allow utf-8 input
\usepackage[T1]{fontenc}    % use 8-bit T1 fonts
\usepackage{hyperref}       % hyperlinks
\usepackage{url}            % simple URL typesetting
\usepackage{booktabs}       % professional-quality tables
\usepackage{amsfonts}       % blackboard math symbols
\usepackage{nicefrac}       % compact symbols for 1/2, etc.
\usepackage{microtype}      % microtypography
\usepackage{xcolor}         % colors
\usepackage{graphicx}
\usepackage[ruled,vlined]{algorithm2e}
\usepackage{makecell}
\usepackage{multirow}
\usepackage{float}
\usepackage{cancel}
\usepackage[compact]{titlesec}
\usepackage{amsmath}
\usepackage{caption}
\usepackage{subfigure}
\usepackage{capt-of}
\usepackage{wrapfig}
\usepackage{array}
\newcolumntype{?}{!{\vrule width 1.5pt}}

\AtBeginDocument{%
  \providecommand\BibTeX{{%
    \normalfont B\kern-0.5em{\scshape i\kern-0.25em b}\kern-0.8em\TeX}}}

\copyrightyear{2023} 
\acmYear{2023} 
\setcopyright{acmlicensed}\acmConference[WSDM '23]{Proceedings of the Sixteenth ACM International Conference on Web Search and Data Mining}{February 27-March 3, 2023}{Singapore, Singapore}
\acmBooktitle{Proceedings of the Sixteenth ACM International Conference on Web Search and Data Mining (WSDM '23), February 27-March 3, 2023, Singapore, Singapore}
\acmPrice{15.00}
\acmDOI{10.1145/3539597.3570403}
\acmISBN{978-1-4503-9407-9/23/02}

\begin{document}

\title{Search Behavior Prediction: A Hypergraph Perspective}

% \author{Yan Han, Edward W Huang, Wenqing Zheng, Nikhil Rao, Zhangyang Wang, Karthik Subbian\\
% \texttt{\{yannhan,ewhuang,nikhilsr,ksubbian\}@amazon.com},
% \texttt{\{w.zheng,atlaswang\}@utexas.com}
% }
\author{Yan Han}
\affiliation{%
  \institution{University of Texas at Austin}
  \streetaddress{}
  \city{Austin, TX}
  \country{USA}
  }
\email{yh9442@utexas.edu}

\author{Edward W Huang}
\affiliation{%
  \institution{Amazon}
  \streetaddress{}
  \city{Palo Alto, CA}
  \country{USA}
  }
\email{ewhuang@amazon.com}

\author{Wenqing Zheng}
\affiliation{%
  \institution{University of Texas at Austin}
  \streetaddress{}
  \city{Austin, TX}
  \country{USA}
  }
\email{w.zheng@utexas.edu}

\author{Nikhil Rao}
\affiliation{%
  \institution{Amazon}
  \streetaddress{}
  \city{Palo Alto, CA}
  \country{USA}
  }
\email{nikhilsr@amazon.com}

\author{Zhangyang Wang}
\affiliation{%
  \institution{University of Texas at Austin}
  \streetaddress{}
  \city{Austin, TX}
  \country{USA}
  }
\email{atlaswang@utexas.edu}

\author{Karthik Subbian}
\affiliation{%
  \institution{Amazon}
  \streetaddress{}
  \city{Palo Alto, CA}
  \country{USA}
  }
\email{ksubbian@amazon.com}

\renewcommand{\shortauthors}{Han, Huang, Zheng, Rao, Wang and Subbian}

\begin{abstract}
    At E-Commerce stores such as Amazon, eBay, and Taobao, the shopping items and the query words that customers use to search for the items form a bipartite graph that captures search behavior. Such a query-item graph can be used to forecast search trends or improve search results. For example, generating query-item associations, which is equivalent to predicting links in the bipartite graph, can yield recommendations that can customize and improve the user search experience. Although the bipartite shopping graphs are straightforward to model search behavior, they suffer from two challenges: 1) The majority of items are sporadically searched and hence have noisy/sparse query associations, leading to a \textit{long-tail} distribution. 2) Infrequent queries are more likely to link to popular items, leading to another hurdle known as \textit{disassortative mixing}.
   
    To address these two challenges, we go beyond the bipartite graph to take a hypergraph perspective, introducing a new paradigm that leverages \underline{auxiliary} information from anonymized customer engagement sessions to assist the \underline{main task} of query-item link prediction. This auxiliary information is available at web scale in the form of search logs. We treat all items appearing in the same customer session as a single hyperedge. The hypothesis is that items in a customer session are unified by a common shopping interest. With these hyperedges, we augment the original bipartite graph into a new \textit{hypergraph}. We develop a \textit{\textbf{D}ual-\textbf{C}hannel \textbf{A}ttention-Based \textbf{H}ypergraph Neural Network} (\textbf{DCAH}), which synergizes information from two potentially noisy sources (original query-item edges and item-item hyperedges). In this way, items on the tail are better connected due to the extra hyperedges, thereby enhancing their link prediction performance. We further integrate DCAH with self-supervised graph pre-training and/or DropEdge training, both of which effectively alleviate disassortative mixing. Extensive experiments on three proprietary E-Commerce datasets show that DCAH yields significant improvements of up to \textbf{24.6\% in mean reciprocal rank (MRR)} and \textbf{48.3\% in recall} compared to GNN-based baselines. Our source code is available at \url{https://github.com/amazon-science/dual-channel-hypergraph-neural-network}.
\end{abstract}





\maketitle

\section{Introduction}
Customers shop online in global E-Commerce stores such as Amazon, eBay, or Taobao for a variety of items. The stores' search engines are responsible for understanding each customer's query intent and then retrieving a list of relevant items. The retrieval processes often use algorithms that leverage a mix of lexical and behavioral signals. Lexical features include the query text and item titles, while behavioral signals include actions that customers take after executing queries, such as clicking on or purchasing a retrieved item.

In this work, we focus on modeling behavioral signals by first representing customer queries and items in the store catalogs as nodes in a bipartite graph. The behavioral signals (\textit{e.g.}, clicks and purchases) serve as edges in the graph by capturing interactions between queries and items. Machine learning models can then learn to predict the query-item edges from this graph in the link prediction task. These predicted behavioral signals can further improve retrieval algorithms and therefore the customer experience.

Graph neural networks (GNNs) \cite{gcn, graphsage} have achieved state-of-the-art results for link prediction \cite{zhang2018link} across a wide range of graph-structured data. Despite strong performance on many types of graphs, they still suffer from two challenges caused by the raw query-item bipartite graph.
\begin{itemize}
    \item First, the success of GNNs hinges on dense and high-quality connections. Unfortunately, in typical bipartite shopping graphs, few items are popular enough to be consistently searched for, such as everyday household items. The majority of items face the cold-start problem \cite{leroy2010cold}, where very few interactions from customers lead to sparse and noisy query-item edges. These items form a less reliable part of the graph, known as the long-tail part, and typical GNNs do not perform well on the long-tail part \cite{leroy2010cold, ding2021zero}.
    \item The second challenge lies in the disassortative mixing commonly observed in our bipartite shopping graphs \cite{adhikari2018mining}: unpopular queries/items are more likely to be connected to popular items/queries. We show through experimental results that disassortative mixing is a major hurdle for query-item predictions. We further demonstrate that disassortative mixing in shopping graphs leads to the over-smoothing \cite{li2018deeper} problem, triggering performance drops in GNN link prediction. More detailed illustrations of why and how disassortative mixing happens in our graph are in Sec~\ref{sec:3.2.2}.
\end{itemize}

\vspace{-0.3em}
We explore the two aforementioned challenges, identifying their causes and proposing solutions. Our solution involves using the wealth of meta-information in the bipartite shopping graph, which is often overlooked. We refer to this meta-information as \textit{auxiliary} information. An example of auxiliary information is anonymized search records. These records are available at web scale but are difficult to model in normal bipartite graphs due to their higher-order relations. However, they can contain information not present in the original pairwise relations. For example, items occurring in the same customer search session may be connected by a common theme that query-item links do not capture. Therefore, such auxiliary information can be leveraged to assist the main task of query-item link prediction. A detailed illustration of the advantages of utilizing the hypergraph to model the auxiliary information is in Sec~\ref{sec:3.2.1}. Both the original bipartite graph and a hypergraph constructed from auxiliary information provide valuable information that the other graph does not. Therefore, we jointly consider both graphs when predicting user search behavior. Our goal is to determine \textit{how to optimally leverage the auxiliary hypergraph to assist the main task of query-item link prediction.}

\begin{figure*}
  \centering
  \includegraphics[width=0.85\linewidth]{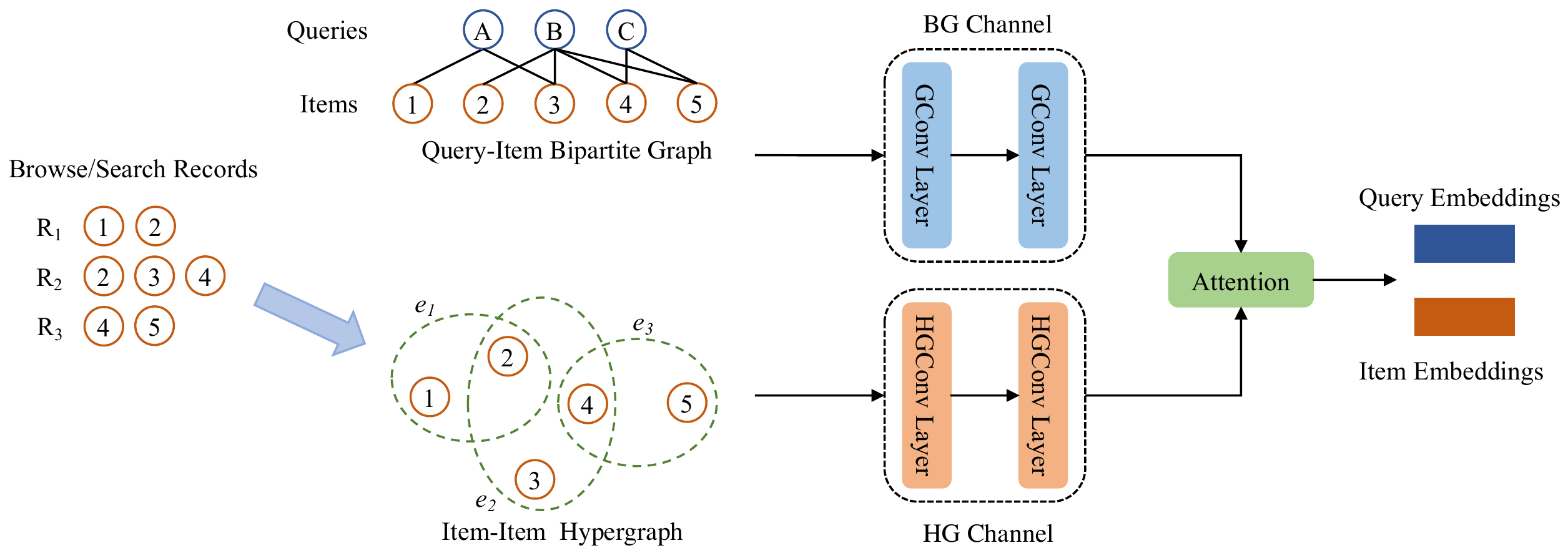}
  \caption{Construction of the hypergraph and the pipeline of the proposed DCAH model.}
  \label{model}
  \vspace{-0.4cm}
\end{figure*}

In this paper, we study the problem of search behavior prediction in the following setting: given two different yet related graphs, a bipartite graph and a hypergraph, we aim to jointly use them to predict query-item links. Our hypothesis is that using both graphs is better than using just one. We conceptually illustrate this unique problem setting in Figure~\ref{model}. 
Taking the E-Commerce problem as an example, we consider two different graphs: a bipartite query-item graph constructed from search logs and a hypergraph constructed from auxiliary information. Although the two graphs share the same node set, their distributions may be different. For instance, everyday items are densely connected nodes in the query-item graph due to their popularity. On the other hand, they may be searched and purchased alone in customer sessions, resulting in few connections in the hypergraph.

To tackle this problem setting, we first propose the \textit{\textbf{D}ual-\textbf{C}hannel \textbf{A}ttention-Based \textbf{H}ypergraph Neural Network}, or \textbf{DCAH}, which can jointly learn better representations from the two input graphs. In particular, we use the bipartite graph to model the pairwise relations while leveraging the hypergraph to model the higher-order relations. To integrate these two relations, we leverage an attention mechanism to learn the optimal weight each relation should have for the downstream task. Furthermore, to improve DCAH's ability to generalize to the long tail and alleviate the over-smoothing problem, we incorporate self-supervised learning and the DropEdge \cite{rong2020dropedge} strategies. Self-supervised learning works well for few-shot generalization \cite{chen2020big}, which can alleviate the fact that the long tail faces label sparsity. Meanwhile, DropEdge relieves the over-smoothing problem \cite{cai2020note, huang2020tackling} introduced by disassortative mixing. We provide a detailed illustration of the two strategies in Sec~\ref{sec:3.4}. In summary, this paper makes the following contributions:
\vspace{-0.4em}
\begin{itemize}
    \item We study the query-item link prediction problem in a special setting: improving the main task of search behavior prediction (link prediction) with the aid of accessible auxiliary information. We use the auxiliary information to create a second hypergraph supplementing the original bipartite graph.
    \item We propose a new framework, DCAH, to tackle our unique problem setting by jointly capturing the information of the bipartite graph and the hypergraph. Additionally, we seamlessly incorporate self-supervised learning to improve DCAH's generalization ability on the original graph's long tail and DropEdge to relieve the over-smoothing problem.
    \item We conduct experiments on three proprietary E-Commerce datasets. Our experimental results show that our approach not only improves the overall link prediction performance, but also generalizes better to the long tail compared to existing off-the-shelf approaches.
\end{itemize}

\section{Related Work}

\vspace{-0.1em}
\subsection{Session-Based Recommendation}
Session-based recommendation (SBR) shares some characteristics with our problem setting. However, though both settings involve user search behavior, they differ in their motivations, solutions, and final goals. Generally, a session is a transaction with multiple purchased items in one shopping event. SBR focuses on next-item prediction by using real-time user behavior \cite{xia2021self}. It can be seen as a form of personalized recommendation. However, we focus on query-item link prediction rather than user-oriented recommendation, and our problem setting does not involve inter-user relationships.

Additionally, initial exploration in SBR mainly focuses on sequence modeling, including traditional Markov decision processes \cite{zimdars2013using} and deep learning models such as recurrent neural networks (RNNs) and convolutional neural networks (CNNs) \cite{tuan20173d}. More recently, GNN has shown promising results in SBR \cite{wang2020beyond, wu2019session}. To capture the complex higher-order item correlations, DHCN \cite{xia2021self} and SHARE \cite{wang2021session} constructed session-induced hypergraphs to model item correlations and user intent in SBR, which achieved superior performance against previous GNN-based models. This also indicates the advantages of hypergraphs over normal graphs when modeling high-order relations.

\vspace{-0.1em}
\subsection{Graphs/Hypergraphs in Recommendation}
Since graphs can naturally model relational data such as item-item, user-user, or user-item interactions, GNN-based models have performed well in recommendation systems \cite{kherad2020recommendation, shaikh2017recommendation}. GCMC \cite{berg2017graph} utilized neural graph autoencoders to reconstruct a user-item rating graph. NGCF \cite{wang2019neural} proposed to construct a user-item bipartite graph and leverage GNNs to model relations between users and items. In social recommendation, relations between users can be exploited with social graphs \cite{fan2019graph, wu2019session}. For example, DiffNet \cite{fan2019graph} adopted GCNs to model the diffusion of user embeddings among their social connections. In addition, inspired by the ability of hypergraphs to model complex higher-order dependencies \cite{feng2019hypergraph, gao2020hypergraph}, recent works have attempted to capture interactions via hypergraph structure and uniform node-hyperedge connections, including HyRec \cite{wang2020next}, DHCF \cite{ji2020dual}, and MHCN \cite{yu2021self}. HyRec attempted to propagate information among multiple items by considering users as hyperedges. DHCF modeled hybrid multi-order correlations between users and items based on the hypergraph structure. MHCN leveraged social relations to create a hypergraph that improves recommendation quality when user-item interaction data is sparse.

\vspace{-0.1em}
\subsection{Contrastive Representation Learning}
Contrastive learning effectively captures consistent feature representations under different views \cite{van2018representation}. It has achieved promising results in various domains, such as visual data representation \cite{peng2021self}, language data understanding \cite{cao2021grammatical}, graph representation learning \cite{qiu2020gcc, you2020graph}, and recommendation systems \cite{yu2021self, xia2022hypergraph}. These contrastive learning approaches use auxiliary signals specific to their data or tasks. Concurrently, HCCF \cite{xia2022hypergraph} jointly captured local and global collaborative relations with a hypergraph-enhanced cross-view contrastive learning architecture. However, their self-supervised learning setting is different from ours, since we augment the original bipartite graph and hypergraph and maximize the agreement between the two generated variants, following the standard graph contrastive learning pipeline \cite{you2020graph}.

\vspace{-0.1em}
\subsection{Disassortative Mixing}
Many scale-free networks in the real world show the tendency where highly connected nodes link with other highly connected nodes, which is defined as assortative mixing. The reverse is also true in some networks, where highly connected nodes are more likely to make links with isolated, less connected nodes, which is defined as disassortative mixing. Biological and technological networks typically show a disassortative mixing pattern. In \cite{hu2009disassortative}, the authors found that online social networks also exhibit a disassortative mixing pattern, and \cite{adhikari2018mining} observed disassortative mixing in the customer interaction networks of E-Commerce stores. Although disassortative mixing has been observed in the E-Commerce domain, to the best of our knowledge, no formal study on how disassortative mixing affects the performance of recommendation models and how to relieve the issue currently exists. In this paper, we reveal how disassortative mixing triggers the over-smoothing problem and utilize two existing off-the-shelf approaches, self-supervised learning and DropEdge \cite{rong2020dropedge}, to mitigate it.

\section{Methodology}
In this section, we first introduce the notations and definitions used throughout the paper, and then we show how the auxiliary information is modeled as a hypergraph. After that, we revisit the two challenges in the query-item graph. Then, we present the DCAH framework. Finally, we discuss how we integrate self-supervised learning and DropEdge to improve the generalization ability of our model and relieve the over-smoothing problem.

\subsection{Notations, Definitions, and Hypergraph Construction}
\paragraph{\textbf{Notations.}} Let $Q = \{q_1, q_2, q_3, \ldots, q_{N_Q}\}$ and $I = \{i_1, i_2, i_3, \ldots, i_{N_I}\}$ denote the sets of queries and items, where $N_Q$ and $N_I$ are the number of queries and the number of items, respectively. $N$ denotes the total number of queries and items. Each browse/search record is represented as a set $\{i_{r,1}, i_{r,2}, i_{r,3}, \ldots, i_{r,m}\}$, where $i_{r,k} \in I (1 \leq k \leq m)$ represents an interaction between an anonymous user and an item within the browse/search record $r$.

\paragraph{\textbf{Hypergraph Definition.}} A hypergraph is defined as $\mathcal{G} = (\mathcal{V}, \mathcal{E})$, where $\mathcal{V}$ is a set of $N$ unique vertices and $\mathcal{E}$ is a set of $M$ hyperedges. Each hyperedge $e \in \mathcal{E}$ connects two or more vertices. The hypergraph can be represented by an incidence matrix $\mathbf{H} \in \mathbb{R}^{N\times M}$, where $H_{ie} = 1$ if the hyperedge $e \in \mathcal{E}$ contains a vertex $v_i \in \mathcal{V}$, otherwise $0$. For each vertex and hyperedge, their degrees $D_{ii}$ and $B_{ee}$ are defined as $D_{ii} = \sum_{e=1}^M H_{ie}$ and $B_{ee} = \sum_{i=1}^N = H_{ie}$, respectively. $\mathbf{D}$ and $\mathbf{B}$ are diagonal matrices. $D_{ii}$ and $B_{ee}$ are the $i$-th and $e$-th diagonal elements of $\mathbf{D}$ and $\mathbf{B}$, respectively.

\paragraph{\textbf{Hypergraph Construction.}} To capture relationships beyond pairwise ones in the browse/search records, we adopt a hypergraph  $\mathcal{G} = (\mathcal{V}, \mathcal{E})$ to represent each record as a hyperedge. Formally, we denote each hyperedge as $\{i_{r,1}, i_{r,2}, i_{r,3}, \ldots, i_{r,m}\} \in \mathcal{E}$ and each item as $i_{r, k} \in \mathcal{V} (1 \leq k \leq m)$. The hypergraph construction process is shown in the left part of Figure~\ref{model}. As illustrated, the original record data is organized as linear sequences where two items, $i_{r, m-1}$ and $i_{r,m}$, are only connected if a user interacted with $i_{r,m-1}$ before $i_{r,m}$. After transforming the record data into a hypergraph, any two items browsed/searched in the same record are connected. This allows us to concretize the many-to-many higher-order relations.

\subsection{Two Challenges in the Query-Item Graph}
In this section, we revisit the two unique challenges of the raw query-item bipartite shopping graphs: the long-tail distribution and disassortative mixing. Specifically, we investigate the node-degree distribution of the query-item graph and show how the auxiliary hypergraph helps relieve the problem. Additionally, we empirically demonstrate why disassortative mixing is a major hurdle for link prediction models.

\subsubsection{Long-Tail Distribution}
\label{sec:3.2.1}
\begin{figure}
  \centering
  \includegraphics[width=0.85\linewidth]{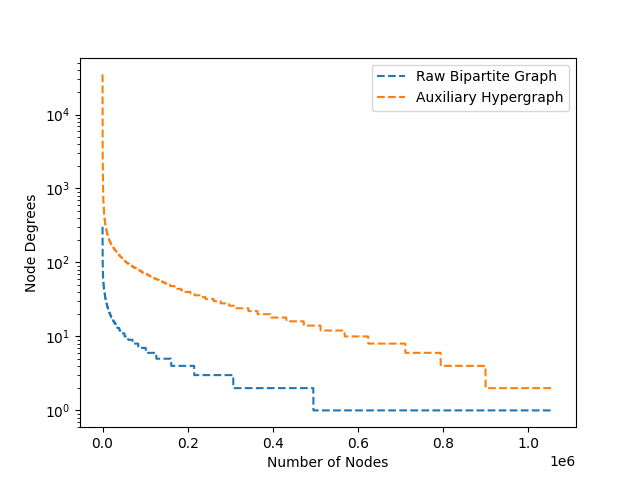}
  \caption{Node-degree distribution of one E-Commerce query-item bipartite graph and the related auxiliary hypergraph.}
  \label{degree}
  \vspace{-0.4cm}
\end{figure}

Many real networks are scale free in nature, i.e, the node degrees follow power-law distributions \cite{broder2000graph}. As a type of real network, query-item graphs also tend to have a similar power distribution pattern. The degree distribution plot of one E-Commerce query-item graph used in our paper is presented in Figure~\ref{degree} (blue line). In the figure, we can find that only a few (less than 10\%) nodes have node degree greater than ten, while the majority of nodes have fewer than ten neighbors. In summary, the degree distribution of the E-Commerce query-item graph follows the long-tail degree distribution. The long-tail degree distribution indicates that while there exist some popular queries/items which connect with many other items/queries, most queries/items connect only to a few items/queries, which is called the long-tail part. In addition, the long-tail part is less reliable since behavioral query-item relations (\textit{e.g.}, from clicks or views) can introduce noise (\textit{e.g.}, misclicks) to the constructed interaction graph \cite{zhao2021heterogeneous}. In this case, existing GNNs fail to perform well on the long-tail part due to data sparsity and noisy interactions. In addition, most GNN-based methods design the message-passing mechanism such that the embedding propagation is only performed with neighbors in the original graph.

As mentioned in the introduction, there exists one type of auxiliary information that we can introduce to alleviate the above issue: anonymized search records. Previous methods, such as the K-Nearest Neighbor (KNN) graph based on node feature similarity \cite{wang2020gcn, tang2015line}, methods based on conventional Markov chains \cite{koren2009matrix, rendle2010factorizing}, and methods based on factorization \cite{liang2016factorization, eirinaki2005web}, have been proposed to model some kinds of auxiliary information in similar problem settings. However, these methods cannot capture higher-order dependencies \cite{wang2022exploiting, wu2021dual} such as those in customer search sessions. Hypergraphs \cite{bretto2013hypergraph}, which generalize the concept of an edge to connect more than two nodes, can organically model complex, higher-order relations. In addition, the auxiliary hypergraph can generally enrich the connectivity of the nodes of the raw bipartite graph. As shown in Figure~\ref{degree} (yellow line), the node degrees are boosted via a huge margin, especially on the long-tail part.

\subsubsection{Disassortative Mixing}
\label{sec:3.2.2}
Since there exist only a few popular queries and items in the query-item graph, we assume it is not uncommon that some popular queries/items may link to unpopular items/queries. To verify this phenomenon, we investigate the popular queries of the graph, and find that the queries with highest degrees tend to be very general queries, such as ``toothbrush'', ``tissue'', ``TV'', etc. Such queries connect to many related items, some of which are unpopular or specialized items, such as \textit{premium bamboo toothbrush}, \textit{baby tissue}, \textit{LED 4K UHD TV} etc. Similarly, for popular everyday household items like \textit{water}, \textit{toilet paper}, etc., due to popularity, many customers may search for them a large number of times. In their search, the customers may generate queries with infrequent terms, such as ``best spring water in the world'' and ``ultra soft toilet paper for infant.'' Based on this observation, we believe that disassortative mixing exists in the E-Commerce query-item graph and conduct the degree assortativity analysis of the graph.

\begin{figure}
  \centering
  \includegraphics[width=0.85\linewidth]{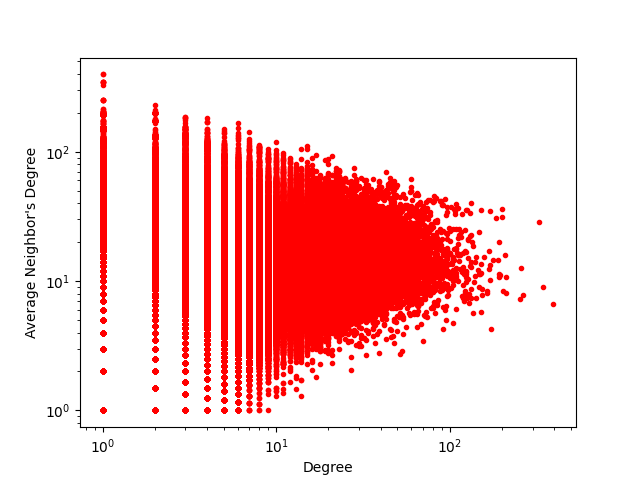}
  \caption{Assortativity plot (degree vs average neighbor’s degree) for one E-Commerce query-item bipartite graph.}
  \label{disassortative}
  \vspace{-0.4cm}
\end{figure}

Degree assortativity, $r \in [-1, 1]$, is a measure of similarity between nodes and their neighbors in terms of degree \cite{newman2003mixing}. Formally, degree assortativity is defined as the Pearson Correlation Coefficient of degrees between all pairs of connected nodes. The value $r=-1$ implies that the network is totally disassortative (negative correlation) and $r=1$ implies that the network is totally assortative (positive correlation). The assortativity plot of one E-Commerce query-item graph used in this paper is shown in Figure~\ref{disassortative}. The query-item bipartite graph is relatively disassortative with $r=-0.07$. In the figure, we observe that the neighbors of high-degree nodes have relatively low degrees, while the neighbors of some low-degree nodes have very high degrees. In summary, the assortativity plot of the query-item graph suggests that unpopular queries/items are more likely to connect to popular items/queries, making the graph disassortative, which agrees with our intuition.

Relative degree evaluates a node's degree compared to its neighbors' degrees \cite{yan2021two}. When all the nodes have the same degree, the relative degree equals 1. On the other hand, relative degree will be low if a node's degree differs significantly from its neighbors' degrees. As a type of disassortative graph, the relative degree of the E-Commerce query-item bipartite graph is relatively low compared to that of assortative graphs. In \cite{yan2021two}, the authors theoretically showed that the low relative degree will trigger the over-smoothing problem, such that all nodes' representations will converge to a stationary point with an increase in the number of layers of graph convolutions. Although over-smoothing is usually discussed in the node classification task, it also acts as a major hurdle in the link prediction task. We show that this is true in our query-item prediction task with our experiment results.

\subsection{A General DCAH Framework}
Figure~\ref{model} sketches the overall framework of DCAH. At a high level, DCAH consists of a bipartite graph channel and a hypergraph channel. The bipartite graph channel captures pairwise query-item relations while the hypergraph channel captures higher-order item-item relations. DCAH uses an attention layer to compute the optimal weighted combination of the channel outputs for the downstream task, query-item prediction.

\paragraph{\textbf{Bipartite Graph Channel and Convolution.}} The original bipartite shopping graph is an undirected graph containing only query-item relations. First, we initialize the item embeddings specific to the bipartite graph channel, $\mathbf{X}_I^{(0)_{BG}}$. For the query node, the raw text of the user query is available to generate node features using Byte-Pair Encodings \cite{heinzerling2018bpemb}, denoted as $\mathbf{X}_Q^{(0)_{BG}}$. The adjacency matrix for this bipartite graph $\mathcal{G}_B$ is defined as $\mathbf{A}\in \mathbb{R}^{N\times N}$. Let $\hat{\mathbf{A}} = \mathbf{A} + \mathbf{I}$, where $\mathbf{I}$ is the identity matrix. $\hat{\mathbf{D}}\in \mathbb{R}^{N\times N}$ is a diagonal degree matrix where $\hat{\mathbf{D}}_{p,p} = \sum_{q=1}^n \hat{\mathbf{A}}_{p,q}$. The bipartite graph convolution is then defined as:
\begin{equation}
    \mathbf{X}^{(l+1)_{BG}} = \hat{\mathbf{D}}^{-1/2}\hat{\mathbf{A}}\hat{\mathbf{D}}^{-1/2}\mathbf{X}^{(l)_{BG}}\Theta_{BG}
\end{equation}
where $\Theta_{BG}$ is the learnable shared parameter of the convolutional layer. In each convolution, the query/item node gathers information from its neighboring item/query nodes. By doing so, the learned $\mathbf{X}$ can capture the pairwise query-item relation information. We then pass $\mathbf{X}^{(0)}$ through $L$ graph convolutional layers to obtain the final output embeddings $\mathbf{X}^{(L)}$. We use $\mathbf{X}_{Q}^{(L)_{BG}}$ and $\mathbf{X}_{I}^{(L)_{BG}}$ to denote query and item embeddings, respectively.

\paragraph{\textbf{Hypergraph Channel and Convolution.}} The hypergraph channel is used to capture the item-level higher-order relations. The primary challenge of defining a convolution operation over the hypergraph is how the embeddings of items are propagated. Similar to the bipartite graph channel, we first initialize the hypergraph-channel-specific item embeddings $\mathbf{X}_I^{(0)_{HG}}$. Following the spectral hypergraph convolution proposed in \cite{feng2019hypergraph}, we define our hypergraph convolution as:
\begin{equation}
    \mathbf{X}_I^{(l+1)_{HG}} = \mathbf{D}^{-1/2}\mathbf{H}\mathbf{W}\mathbf{B}^{-1}\mathbf{H}^T\mathbf{D}^{-1/2}\mathbf{X}_I^{(l)_{HG}}\Theta_{HG}
\end{equation}
Here, $\mathbf{W}$ is the hyperedge weight matrix, where we assign all hyperedges the weight of $1$, and $\Theta_{HG}$ is the learnable shared parameter of the convolutional layer. The hypergraph convolution can be viewed as a two-stage refinement performing a node-hyperedge-node feature transformation on the hypergraph structure. The multiplication of $\mathbf{H}^T$ implements the information aggregation from nodes to hyperedges. Pre-multiplying $\mathbf{H}$ can be viewed as aggregating information from hyperedges to nodes. Note that $\mathbf{D}$ and $\mathbf{B}$ play the role of normalization. After passing $\mathbf{X}_I^{(0)_{HG}}$ through $L$ hypergraph convolutional layers, we obtain the final item embeddings $\mathbf{X}_I^{(L)_{HG}}$.

\paragraph{\textbf{Inter-Channel Aggregation.}} With the dual-channel structure, we generate two groups of item embeddings, $\mathbf{X}_{I}^{(L)_{BG}}$ and $\mathbf{X}_I^{(L)_{HG}}$, via the bipartite graph and hypergraph channels, respectively. Since both groups of item embeddings will contain some valuable information, we add an additional attention layer to learn the optimal weighted combination of the two groups of embeddings for our downstream task. This strategy will improve robustness when fusing the two groups' embeddings. Specifically, we first transform both channels' embeddings through a nonlinear transformation (\textit{e.g.}, a single-layer MLP). Then, we measure the importance of a channel-specific embedding as the similarity between the transformed embedding and a channel-level attention vector $\mathbf{q}$. Finally, we average the importance of all channel-specific item embeddings. We can interpret this as the importance of each channel, denoted as $w_{\phi_c}$.
\begin{equation}
    w_{\phi_c} = \frac{1}{|\mathcal{V}|} \sum_{i\in \mathcal{V}}\mathbf{q}^T\cdot \tanh{(\mathbf{W}\cdot\mathbf{x}_i^{\phi_c}+\mathbf{b})}, \ {\phi_c} \in \{BG, HG\} 
\end{equation}
Here, $\mathbf{W}$ is the weight matrix, $\mathbf{b}$ is the bias vector, $\mathbf{x}_i^{\phi_c}$ is the channel-level embedding, $\tanh(\cdot)$ is the activation function, $\mathbf{q}$ is the channel-level attention vector, and $\mathcal{V}$ is the set of the item nodes. Note that, for the sake of meaningful comparison, all the above parameters are shared across all channels and channel-specific embeddings. After obtaining the importance of each channel, we normalize them with the softmax function. The weight of channel $\phi_c$, denoted as $\beta_{\phi_c}$, can be obtained by normalizing the above importance of all channels using the softmax operator:
\begin{equation}
    \beta_{\phi_c} = \frac{\exp{w_{\phi_c}}}{\sum_{\phi_c} \exp{w_{\phi_c}}}, \ {\phi_c} \in \{BG, HG\} 
\end{equation}
$\beta_{\phi_c}$ can be interpreted as the contribution of channel $\phi_c$ for a specific task, with a higher $\beta_{\phi_c}$ indicating a higher importance for channel $\phi_c$. With the learned weights as coefficients, we can fuse the two channel-specific embeddings to obtain the final item embedding $\mathbf{X}_{I}^{L}$ as follows:
\begin{equation}
    \mathbf{X}_{I}^{L} = \beta_{\phi_{BG}}\cdot \mathbf{X}_{I}^{(L)_{BG}} + \beta_{\phi_{HG}}\cdot \mathbf{X}_{I}^{(L)_{HG}}
\end{equation}

\subsection{Enhancing DCAH with Self-Supervised Learning and DropEdge}
\label{sec:3.4}
Although the auxiliary hypergraph can augment the original bipartite graph, data sparsity, the long-tail issue, and disassortative mixing still exist, which might impede the generalization ability of our model. Inspired by prior studies of self-supervised learning and DropEdge on graphs, we seamlessly integrate these two simple but effective training tricks into DCAH.

\paragraph{\textbf{Self-Supervised Learning (SSL).}} In this work, we follow the graph self-supervised learning framework proposed in \cite{you2020graph}, which consists of the four major components: 1) Graph data augmentation; 2) GNN-based encoder; 3) Projection head; 4) Contrastive loss function. It should be noted that we made some modifications to components 1, 2, and 4. Specifically, for the graph data augmentation, the original framework only has one input graph, while we consider two input graphs, the bipartite graph $\mathcal{G}_B$ and the hypergraph $\mathcal{G}_H$. Although the hypergraph is different from the bipartite graph, the representation formats are similar. Therefore, both given graphs <$\mathcal{G}_B$, $\mathcal{G}_H$> can undergo graph data augmentations proposed in \cite{you2020graph} to obtain two correlated views, <$\hat{\mathcal{G}}_{B_i}$, $\hat{\mathcal{G}}_{H_i}$>, <$\hat{\mathcal{G}}_{B_j}$, $\hat{\mathcal{G}}_{H_j}$>, as a positive pair. Then, our model DCAH extracts node-level embeddings $\mathbf{h}_i$ and $\mathbf{h}_j$ for two augmented graph pairs <$\hat{\mathcal{G}}_{B_i}$, $\hat{\mathcal{G}}_{H_i}$> and <$\hat{\mathcal{G}}_{B_j}$, $\hat{\mathcal{G}}_{H_j}$>, respectively. After that, as advocated in \cite{chen2020simple}, a projection head in the form of a two-layer perceptron (MLP) is applied to obtain $z_i$ and $z_j$. Finally, a node-level contrastive loss $\mathcal{L}(\cdot)$ is utilized to enforce maximizing the consistency between positive pairs $z_i$ , $z_j$ compared with negative pairs:
\begin{equation}
    \mathcal{L} = -\log\frac{\exp(sim(z_{n,i}, z_{n,j})/\tau)}{\sum_{n'=1, n' \neq n}^{N} \exp(sim(z_{n,i}, z_{n',j})/\tau)}
\end{equation}
where $sim(z_{n,i}, z_{n,j})=z_{n,i}^Tz_{n,j}/||z_{n,i}||||z_{n,j}||$, $z_{n,i}$, $\tau$ is the temperature parameter and $N$ is number of nodes.

In summary, our contrastive learning component maximizes the agreement between two variants of two graph augmentations. Furthermore, the setting is different from that of cross-channel views contrastive learning \cite{xia2021self} and global-local views contrastive learning \cite{xia2022hypergraph}.

\paragraph{\textbf{DropEdge.}} As mentioned in Sec~\ref{sec:3.2.2}, the disassortative mixing of the query-item graph triggers the over-smoothing problem, which acts as a major hurdle to the models' performance on the query-item predictions. Previous literature \cite{rong2020dropedge, cai2020note, huang2020tackling} have theoretically and empirically shown that the DropEdge \cite{rong2020dropedge} can help relieve over-smoothing. Because DropEdge randomly removes a certain number of edges from the input graph at each training epoch, it can be easily applied to hypergaphs, functioning as a data augmentation mechanism. Therefore, it can be flexibly incorporated into our model DCAH, similar to self-supervised learning.

\section{Evaluation}
Our experiments aim to answer the following research questions:
\begin{itemize}
\item \textbf{RQ1}: What is the performance of our DCAH compared to baselines?
\item \textbf{RQ2}: How does the auxiliary hypergraph contribute to relieving the long-tail issue of the raw bipartite graph?
\item \textbf{RQ3}: How do SSL and DropEdge perform in alleviating the over-smoothing problem triggered by disassortative mixing?
\end{itemize}

\subsection{Experimental Settings}
\subsubsection{Dataset Overview.} We evaluate the proposed framework on three proprietary E-Commerce datasets of queries and items, which are sampled from three different market locales of this E-Commerce platform. They can all be naturally constructed as bipartite shopping graphs. The query-item associations are sampled so as not to reveal raw traffic distributions. We also create three related auxiliary item-item hypergraphs from the three locales' corresponding anonymized customer engagement sessions, where all customer-identifiable information has been properly anonymized. The full dataset statistics are summarized in Table~\ref{stat}. In the table, the density indicates that all the three E-Commerce datasets are generally sparse, among which E-Commerce 3 is the sparsest. As for the degree assortativity, E-Commerce 3 is more disassortative than E-Commerce 1 and 2.

\begin{table}[h]
    \centering
    \caption{Dataset statistics.}
    \vspace{-0.4cm}
    \resizebox{0.45\textwidth}{!}{
    \begin{tabular}{c|c|c|c} 
    \toprule[1.5pt]
      & E-Commerce 1 & E-Commerce 2 & E-Commerce 3 \\ 
      \midrule
    Num. of Nodes & 1,181,247 & 1,059,963 & 1,375,842 \\ 
    \midrule
    Num. of Edges & 1,783,333 & 1,634,402 & 2,148,822 \\
    \midrule
    Num. of Auxiliary HyperEdges & 329,852 & 238,342 & 388,042 \\
    \midrule
    Degree Assortativity & 0.132 & 0.089 & -0.07 \\
    \midrule
    Density & $1.28\times e^{-6}$ & $1.45\times e^{-6}$ & $1.14\times e^{-6}$ \\
    
    \bottomrule[1.5pt]
    \end{tabular}}
    \vspace{-0.3cm}
    \label{stat}
\end{table}

\subsubsection{Data Splits.} We create training, validation, and test splits of the data to specifically study tail and cold-start nodes. Because we have a bipartite graph and a hypergraph, we manually split the nodes into four parts, which we refer to as the \textit{head, tail1, tail2}, and \textit{isolation} parts.

We create the \textit{head} part by first selecting nodes that are in the top $20\%$ highest-degree nodes of both the bipartite graph and the hypergraph, then inducing a subgraph with the selected nodes. The \textit{tail1} part induces a subgraph on the nodes shared between the top $20\%$ highest-degree nodes of bipartite graph and the top $20\%-60\%$ highest-degree nodes of hypergraph. Similarly, the \textit{tail2} part induces a subgraph on the nodes shared between the top $20\%$ highest-degree nodes of hypergraph and the top $20\%-60\%$ highest-degree nodes of the bipartite graph. The \textit{isolation} part induces a subgraph on the nodes in the bottom $40\%$ highest-degree nodes of both the bipartite graph and the hypergraph.

For each part, we randomly sample $20\%$ of the edges for testing and $10\%$ for validation. The remaining edges are used for training. In summary, our data split is 70\% training, 10\% validation, and 20\% testing. Within the test set, each of the four parts, \textit{head, tail1, tail2}, and \textit{isolation}, counts as $25\%$. In addition, since the downstream task is link prediction, we must sample negative edges. For each positive edge in the original graph, containing (\textit{source}, \textit{destination}) nodes, we corrupt the edge by replacing its \textit{source} or \textit{destination} with 100 randomly sampled negative edges (50 for \textit{source} and 50 for \textit{destination}), while ensuring the resulting negative edges are not already present in the original graph. We ensure no overlap between the positive and negative edges across all split parts.

\subsubsection{Evaluation Metrics.} We use \textit{Recall@N} and Mean Reciprocal Rank (\textit{MRR}) as the evaluation metrics, which are widely used in link prediction tasks \cite{hu2020open}. $N$ is the number of the top predictions. In our experiments, we set $N$ to be the number of the true positive edges of the testing part. \textit{Recall@N} measures whether the actual edge is on the top-$N$ predictions. Thus, it denotes the proportion of the number of predictions containing the actual edge among all positive edges.
% as follows:
% \begin{equation}
%     \textit{Recall@$N$} = \frac{n_{hit}}{N},
% \end{equation}
% where $n_{hit}$ is the number of predictions having the actual edge. 
\textit{MRR} measures the ranking performance of the model as an average of reciprocal ranks of the actual edge within the prediction.
% as follows:
% \begin{equation}
%     \textit{MRR} = \frac{1}{N} \sum_{c\in N} \frac{1}{rank(c)},
% \end{equation}
% where $c$ is the actual edge. 
Both evaluation metrics are appropriate to measure the model's performance, but \textit{Recall@N} is more related to the ability to find unseen edges. Unseen edges are scarce in the tail and isolation parts and are therefore more important to predict. Thus, \textit{Recall@N} is more important than \textit{MRR}, especially under the cold-start scenario. We report the mean and standard deviation of \textit{Recall@N} and \textit{MRR} after $10$ runs. 

\subsubsection{Baselines.} Due to the uniqueness of our problem setting, the meaningful baselines are limited. In our experiments, we compare our DCAH with two commonly used baselines: GCN \cite{gcn}, and HyperGCN \cite{feng2019hypergraph}. And for fair comparison, the bipartite graph channel and the hypergraph channel use the GCN and HyperGCN, respectively. Besides, we also adapt the SSL and/or DropEdge strategies to the GCN and HyperGCN as additional baselines.

\subsubsection{Hyperparamter Settings.} We use Adam optimizer with the learning rate of $1e-3$. The hidden state dimension is 64. For both the bipartite graph and hypergraph channels, we stack two propagation layers. The batch size and dropout ratio are set to 1024 and 0.25, respectively. The temperature parameter $\tau$ is set to 0.1 to control the strength of gradients in our contrastive learning.

\begin{table}[t]
    \centering
    \caption{Performance comparison on the E-Commerce 1 in terms of \textit{MRR} and \textit{Recall}. The best results are bold.}
    \vspace{-0.4cm}
    \resizebox{0.45\textwidth}{!}{
    \begin{tabular}{l|cc|cc|cc|cc}
    \toprule[1.5pt]
    \multirow{2.5}{*}{\bf Models} & \multicolumn{2}{c}{\bf Head} & \multicolumn{2}{|c}{\bf Tail1} & \multicolumn{2}{|c}{\bf Tail2 } & \multicolumn{2}{|c}{\bf Isolation}  \\ 
    \cmidrule(r){2-9}
    & MRR & Recall@N & MRR & Recall@N & MRR & Recall@N & MRR & Recall@N \\ 
    \midrule[1.5pt]
    GCN & $\underset{\scriptscriptstyle{\pm 3.10}}{14.68}$ & $\underset{\scriptscriptstyle{\pm 1.30}}{6.35}$ & $\underset{\scriptscriptstyle{\pm 2.66}}{13.83}$ & $\underset{\scriptscriptstyle{\pm 1.36}}{4.86}$ & $\underset{\scriptscriptstyle{\pm 1.60}}{15.70}$ & $\underset{\scriptscriptstyle{\pm 0.85}}{3.44}$ & $\underset{\scriptscriptstyle{\pm 6.48}}{25.35}$ & $\underset{\scriptscriptstyle{\pm 1.16}}{3.77}$ \\
     \cmidrule(r){1-1}
     GCN+DropEdge & $\underset{\scriptscriptstyle{\pm 2.37}}{23.24}$ & $\underset{\scriptscriptstyle{\pm 1.74}}{14.11}$ & $\underset{\scriptscriptstyle{\pm 2.07}}{23.22}$ & $\underset{\scriptscriptstyle{\pm 1.51}}{14.73}$ & $\underset{\scriptscriptstyle{\pm 5.15}}{20.95}$ & $\underset{\scriptscriptstyle{\pm 3.44}}{11.02}$ & $\underset{\scriptscriptstyle{\pm 6.52}}{34.59}$ & $\underset{\scriptscriptstyle{\pm 1.85}}{18.71}$ \\
     \cmidrule(r){1-1}
     GCN+SSL &  $\underset{\scriptscriptstyle{\pm 2.89}}{58.72}$ & $\underset{\scriptscriptstyle{\pm 2.37}}{34.24}$ & $\underset{\scriptscriptstyle{\pm 2.65}}{58.05}$ & $\underset{\scriptscriptstyle{\pm 2.01}}{33.85}$ & $\underset{\scriptscriptstyle{\pm 2.46}}{50.48}$ & $\underset{\scriptscriptstyle{\pm 2.38}}{31.69}$ & $\underset{\scriptscriptstyle{\pm 0.79}}{51.10}$ & $\underset{\scriptscriptstyle{\pm 0.41}}{31.48}$ \\
     \cmidrule(r){1-1}
     GCN+SSL+DropEdge & $\underset{\scriptscriptstyle{\pm 2.04}}{56.68}$ & $\underset{\scriptscriptstyle{\pm 2.33}}{33.95}$ & $\underset{\scriptscriptstyle{\pm 2.25}}{56.74}$ & $\underset{\scriptscriptstyle{\pm 1.50}}{31.56}$ & $\underset{\scriptscriptstyle{\pm 1.59}}{52.20}$ & $\underset{\scriptscriptstyle{\pm 2.15}}{33.43}$ & $\underset{\scriptscriptstyle{\pm 3.34}}{55.74}$ & $\underset{\scriptscriptstyle{\pm 0.44}}{32.87}$ \\
     \cmidrule(r){1-1}
     HyperGCN &  $\underset{\scriptscriptstyle{\pm 0.84}}{7.47}$ & $\underset{\scriptscriptstyle{\pm 0.79}}{2.10}$ & $\underset{\scriptscriptstyle{\pm 1.00}}{10.77}$ & $\underset{\scriptscriptstyle{\pm 0.71}}{2.36}$ & $\underset{\scriptscriptstyle{\pm 1.89}}{14.26}$ & $\underset{\scriptscriptstyle{\pm 2.14}}{3.34}$ & $\underset{\scriptscriptstyle{\pm 3.99}}{26.15}$ & $\underset{\scriptscriptstyle{\pm 0.95}}{4.08}$ \\
     \cmidrule(r){1-1}
     HyperGCN+DropEdge &  $\underset{\scriptscriptstyle{\pm 1.68}}{12.48}$ & $\underset{\scriptscriptstyle{\pm 4.89}}{10.45}$ & $\underset{\scriptscriptstyle{\pm 1.90}}{16.60}$ & $\underset{\scriptscriptstyle{\pm 4.25}}{9.73}$ & $\underset{\scriptscriptstyle{\pm 4.08}}{21.02}$ & $\underset{\scriptscriptstyle{\pm 3.06}}{10.28}$ & $\underset{\scriptscriptstyle{\pm 4.16}}{24.39}$ & $\underset{\scriptscriptstyle{\pm 5.21}}{12.68}$ \\
     \cmidrule(r){1-1}
     HyperGCN+SSL &  $\underset{\scriptscriptstyle{\pm 1.05}}{40.23}$ & $\underset{\scriptscriptstyle{\pm 1.62}}{30.07}$ & $\underset{\scriptscriptstyle{\pm 0.53}}{36.25}$ & $\underset{\scriptscriptstyle{\pm 0.74}}{27.45}$ & $\underset{\scriptscriptstyle{\pm 3.00}}{57.13}$ & $\underset{\scriptscriptstyle{\pm 1.44}}{38.63}$ & $\underset{\scriptscriptstyle{\pm 3.80}}{58.80}$ & $\underset{\scriptscriptstyle{\pm 0.25}}{38.73}$ \\
     \cmidrule(r){1-1}
     HyperGCN+SSL+DropEdge &  $\underset{\scriptscriptstyle{\pm 0.16}}{41.22}$ & $\underset{\scriptscriptstyle{\pm 1.01}}{31.01}$ & $\underset{\scriptscriptstyle{\pm 0.29}}{35.67}$ & $\underset{\scriptscriptstyle{\pm 0.92}}{28.29}$ & $\underset{\scriptscriptstyle{\pm 0.27}}{62.18}$ & $\underset{\scriptscriptstyle{\pm 0.51}}{44.39}$ & $\underset{\scriptscriptstyle{\pm 6.52}}{62.33}$ & $\underset{\scriptscriptstyle{\pm 1.48}}{44.46}$ \\
     \cmidrule(r){1-1}
     {\bf DCAH} & $\underset{\scriptscriptstyle{\pm 4.36}}{16.06}$ & $\underset{\scriptscriptstyle{\pm 1.62}}{6.40}$ & $\underset{\scriptscriptstyle{\pm 3.83}}{15.37}$ & $\underset{\scriptscriptstyle{\pm 1.23}}{5.09}$ & $\underset{\scriptscriptstyle{\pm 4.09}}{16.25}$ & $\underset{\scriptscriptstyle{\pm 0.94}}{4.02}$ & $\underset{\scriptscriptstyle{\pm 9.54}}{30.10}$ & $\underset{\scriptscriptstyle{\pm 2.11}}{5.10}$ \\
     \cmidrule(r){1-1}
     {\bf DCAH}+DropEdge & $\underset{\scriptscriptstyle{\pm 2.97}}{24.26}$ & $\underset{\scriptscriptstyle{\pm 2.40}}{15.83}$ & $\underset{\scriptscriptstyle{\pm 2.75}}{22.96}$ & $\underset{\scriptscriptstyle{\pm 2.19}}{15.35}$ & $\underset{\scriptscriptstyle{\pm 4.31}}{23.81}$ & $\underset{\scriptscriptstyle{\pm 2.99}}{17.25}$ & $\underset{\scriptscriptstyle{\pm 6.32}}{34.47}$ & $\underset{\scriptscriptstyle{\pm 1.96}}{17.84}$ \\
     \cmidrule(r){1-1}
     {\bf DCAH}+SSL & $\underset{\scriptscriptstyle{\pm 3.55}}{59.47}$ & $\underset{\scriptscriptstyle{\pm 2.29}}{35.56}$ & $\underset{\scriptscriptstyle{\pm 3.30}}{58.03}$ & $\underset{\scriptscriptstyle{\pm 2.05}}{34.47}$ & $\underset{\scriptscriptstyle{\pm 1.05}}{60.88}$ & $\underset{\scriptscriptstyle{\pm 2.17}}{42.69}$ & $\underset{\scriptscriptstyle{\pm 3.60}}{64.00}$ & $\underset{\scriptscriptstyle{\pm 0.92}}{43.78}$ \\
     \cmidrule(r){1-1}
     {\bf DCAH}+SSL+DropEdge & $\underset{\scriptscriptstyle{\pm 3.21}}{\bf 60.96}$ & $\underset{\scriptscriptstyle{\pm 2.32}}{\bf 36.67}$ & $\underset{\scriptscriptstyle{\pm 2.78}}{\bf 59.59}$ & $\underset{\scriptscriptstyle{\pm 2.44}}{\bf 36.32}$ & $\underset{\scriptscriptstyle{\pm 3.24}}{\bf 63.79}$ & $\underset{\scriptscriptstyle{\pm 2.83}}{\bf 45.85}$ & $\underset{\scriptscriptstyle{\pm 3.10}}{\bf 65.70}$ & $\underset{\scriptscriptstyle{\pm 0.68}}{\bf 45.80}$ \\
    \bottomrule[1.5pt]
    \end{tabular}}
    \vspace{-0.3cm}
    \label{tab2}
    \end{table}

\begin{table}[t]
    \centering
    \caption{Performance comparison on the E-Commerce 2 in terms of \textit{MRR} and \textit{Recall}. The best results are bold.}
    \vspace{-0.4cm}
    \resizebox{0.45\textwidth}{!}{
    \begin{tabular}{l|cc|cc|cc|cc}
    \toprule[1.5pt]
    \multirow{2.5}{*}{\bf Models} & \multicolumn{2}{c}{\bf Head} & \multicolumn{2}{|c}{\bf Tail1} & \multicolumn{2}{|c}{\bf Tail2 } & \multicolumn{2}{|c}{\bf Isolation}  \\ 
    \cmidrule(r){2-9}
    & MRR & Recall@N & MRR & Recall@N & MRR & Recall@N & MRR & Recall@N \\ 
    \midrule[1.5pt]
    GCN & $\underset{\scriptscriptstyle{\pm 2.50}}{14.88}$ & $\underset{\scriptscriptstyle{\pm 1.16}}{5.69}$ & $\underset{\scriptscriptstyle{\pm 2.58}}{15.49}$ & $\underset{\scriptscriptstyle{\pm 0.96}}{4.73}$ & $\underset{\scriptscriptstyle{\pm 2.79}}{14.73}$ & $\underset{\scriptscriptstyle{\pm 0.65}}{3.53}$ & $\underset{\scriptscriptstyle{\pm 5.59}}{24.33}$ & $\underset{\scriptscriptstyle{\pm 0.87}}{4.05}$ \\
     \cmidrule(r){1-1}
     GCN+DropEdge & $\underset{\scriptscriptstyle{\pm 2.26}}{23.26}$ & $\underset{\scriptscriptstyle{\pm 1.12}}{13.91}$ & $\underset{\scriptscriptstyle{\pm 2.46}}{24.77}$ & $\underset{\scriptscriptstyle{\pm 0.76}}{12.01}$ & $\underset{\scriptscriptstyle{\pm 5.83}}{24.25}$ & $\underset{\scriptscriptstyle{\pm 0.81}}{10.31}$ & $\underset{\scriptscriptstyle{\pm 7.47}}{39.36}$ & $\underset{\scriptscriptstyle{\pm 0.74}}{12.98}$ \\
     \cmidrule(r){1-1}
     GCN+SSL & $\underset{\scriptscriptstyle{\pm 2.80}}{56.44}$ & $\underset{\scriptscriptstyle{\pm 1.98}}{33.98}$ & $\underset{\scriptscriptstyle{\pm 2.78}}{58.01}$ & $\underset{\scriptscriptstyle{\pm 1.84}}{35.37}$ & $\underset{\scriptscriptstyle{\pm 4.46}}{58.81}$ & $\underset{\scriptscriptstyle{\pm 2.97}}{35.09}$ & $\underset{\scriptscriptstyle{\pm 4.32}}{64.86}$ & $\underset{\scriptscriptstyle{\pm 0.52}}{36.61}$ \\
     \cmidrule(r){1-1}
     GCN+SSL+DropEdge & $\underset{\scriptscriptstyle{\pm 1.74}}{57.86}$ & $\underset{\scriptscriptstyle{\pm 1.59}}{34.20}$ & $\underset{\scriptscriptstyle{\pm 2.70}}{57.98}$ & $\underset{\scriptscriptstyle{\pm 1.30}}{36.91}$ & $\underset{\scriptscriptstyle{\pm 4.33}}{59.55}$ & $\underset{\scriptscriptstyle{\pm 2.80}}{35.88}$ & $\underset{\scriptscriptstyle{\pm 3.45}}{65.19}$ & $\underset{\scriptscriptstyle{\pm 0.45}}{36.90}$ \\
     \cmidrule(r){1-1}
     HyperGCN & $\underset{\scriptscriptstyle{\pm 0.94}}{10.36}$ & $\underset{\scriptscriptstyle{\pm 0.76}}{2.76}$ & $\underset{\scriptscriptstyle{\pm 1.08}}{14.70}$ & $\underset{\scriptscriptstyle{\pm 0.59}}{3.21}$ & $\underset{\scriptscriptstyle{\pm 2.19}}{19.64}$ & $\underset{\scriptscriptstyle{\pm 0.49}}{3.30}$ & $\underset{\scriptscriptstyle{\pm 5.92}}{39.14}$ & $\underset{\scriptscriptstyle{\pm 1.89}}{6.63}$ \\
     \cmidrule(r){1-1}
     HyperGCN+DropEdge & $\underset{\scriptscriptstyle{\pm 2.39}}{27.48}$ & $\underset{\scriptscriptstyle{\pm 2.34}}{7.02}$ & $\underset{\scriptscriptstyle{\pm 2.10}}{32.59}$ & $\underset{\scriptscriptstyle{\pm 3.39}}{10.40}$ & $\underset{\scriptscriptstyle{\pm 4.03}}{41.76}$ & $\underset{\scriptscriptstyle{\pm 6.04}}{17.50}$ & $\underset{\scriptscriptstyle{\pm 8.64}}{53.14}$ & $\underset{\scriptscriptstyle{\pm 7.69}}{20.98}$ \\
     \cmidrule(r){1-1}
     HyperGCN+SSL & $\underset{\scriptscriptstyle{\pm 0.99}}{52.68}$ & $\underset{\scriptscriptstyle{\pm 1.34}}{35.49}$ & $\underset{\scriptscriptstyle{\pm 0.61}}{51.34}$ & $\underset{\scriptscriptstyle{\pm 0.72}}{37.60}$ & $\underset{\scriptscriptstyle{\pm 0.65}}{72.56}$ & $\underset{\scriptscriptstyle{\pm 1.19}}{52.67}$ & $\underset{\scriptscriptstyle{\pm 5.26}}{78.82}$ & $\underset{\scriptscriptstyle{\pm 1.40}}{55.68}$ \\
     \cmidrule(r){1-1}
     HyperGCN+SSL+DropEdge & $\underset{\scriptscriptstyle{\pm 0.37}}{56.26}$ & $\underset{\scriptscriptstyle{\pm 1.23}}{37.53}$ & $\underset{\scriptscriptstyle{\pm 0.16}}{53.03}$ & $\underset{\scriptscriptstyle{\pm 0.83}}{36.69}$ & $\underset{\scriptscriptstyle{\pm 0.49}}{75.10}$ & $\underset{\scriptscriptstyle{\pm 1.27}}{54.47}$ & $\underset{\scriptscriptstyle{\pm 5.70}}{80.73}$ & $\underset{\scriptscriptstyle{\pm 1.44}}{57.26}$ \\
     \cmidrule(r){1-1} 
     {\bf DCAH} & $\underset{\scriptscriptstyle{\pm 5.93}}{17.22}$ & $\underset{\scriptscriptstyle{\pm 1.83}}{6.01}$ & $\underset{\scriptscriptstyle{\pm 5.24}}{17.85}$ & $\underset{\scriptscriptstyle{\pm 1.53}}{5.00}$ & $\underset{\scriptscriptstyle{\pm 2.85}}{17.97}$ & $\underset{\scriptscriptstyle{\pm 0.88}}{3.72}$ & $\underset{\scriptscriptstyle{\pm 6.00}}{36.69}$ & $\underset{\scriptscriptstyle{\pm 1.61}}{4.68}$ \\
     \cmidrule(r){1-1}
     {\bf DCAH}+DropEdge & $\underset{\scriptscriptstyle{\pm 1.02}}{28.58}$ & $\underset{\scriptscriptstyle{\pm 2.02}}{15.25}$ & $\underset{\scriptscriptstyle{\pm 4.35}}{34.43}$ & $\underset{\scriptscriptstyle{\pm 1.16}}{15.16}$ & $\underset{\scriptscriptstyle{\pm 2.66}}{44.73}$ & $\underset{\scriptscriptstyle{\pm 1.50}}{13.73}$ & $\underset{\scriptscriptstyle{\pm 5.65}}{54.99}$ & $\underset{\scriptscriptstyle{\pm 1.17}}{12.53}$ \\
     \cmidrule(r){1-1}
     {\bf DCAH}+SSL & $\underset{\scriptscriptstyle{\pm 0.82}}{69.53}$ & $\underset{\scriptscriptstyle{\pm 1.79}}{46.50}$ & $\underset{\scriptscriptstyle{\pm 0.78}}{70.80}$ & $\underset{\scriptscriptstyle{\pm 1.84}}{47.85}$ & $\underset{\scriptscriptstyle{\pm 0.49}}{82.48}$ & $\underset{\scriptscriptstyle{\pm 1.71}}{57.46}$ & $\underset{\scriptscriptstyle{\pm 0.92}}{81.14}$ & $\underset{\scriptscriptstyle{\pm 0.59}}{58.13}$ \\
     \cmidrule(r){1-1}
     {\bf DCAH}+SSL+DropEdge & $\underset{\scriptscriptstyle{\pm 0.95}}{\bf 72.38}$ & $\underset{\scriptscriptstyle{\pm 2.74}}{\bf 47.59}$ & $\underset{\scriptscriptstyle{\pm 0.97}}{\bf 73.80}$ & $\underset{\scriptscriptstyle{\pm 1.67}}{\bf 48.78}$ & $\underset{\scriptscriptstyle{\pm 0.55}}{\bf 84.46}$ & $\underset{\scriptscriptstyle{\pm 1.43}}{\bf 59.56}$ & $\underset{\scriptscriptstyle{\pm 0.72}}{\bf 83.43}$ &
     $\underset{\scriptscriptstyle{\pm 0.67}}{\bf 59.75}$ \\
    \bottomrule[1.5pt]
    \end{tabular}}
    \vspace{-0.3cm}
    \label{tab3}
    \end{table}

\begin{table}[t]
    \centering
    \caption{Performance comparison on the E-Commerce 3 in terms of \textit{MRR} and \textit{Recall}. The best results are bold.} 
    \vspace{-0.4cm}
    \resizebox{0.45\textwidth}{!}{
    \begin{tabular}{l|cc|cc|cc|cc}
    \toprule[1.5pt]
    \multirow{2.5}{*}{\bf Models} & \multicolumn{2}{c}{\bf Head} & \multicolumn{2}{|c}{\bf Tail1} & \multicolumn{2}{|c}{\bf Tail2 } & \multicolumn{2}{|c}{\bf Isolation}  \\ 
    \cmidrule(r){2-9}
    & MRR & Recall@N & MRR & Recall@N & MRR & Recall@N & MRR & Recall@N \\ 
    \midrule[1.5pt]
  GCN & $\underset{\scriptscriptstyle{\pm 3.48}}{15.49}$ & $\underset{\scriptscriptstyle{\pm 1.22}}{7.28}$ & $\underset{\scriptscriptstyle{\pm 1.50}}{13.44}$ & $\underset{\scriptscriptstyle{\pm 0.63}}{4.08}$ & $\underset{\scriptscriptstyle{\pm 2.11}}{17.04}$ & $\underset{\scriptscriptstyle{\pm 0.40}}{3.38}$ & $\underset{\scriptscriptstyle{\pm 5.55}}{33.93}$ & $\underset{\scriptscriptstyle{\pm 0.89}}{4.12}$ \\
     \cmidrule(r){1-1}
     GCN+DropEdge & $\underset{\scriptscriptstyle{\pm 3.16}}{32.33}$ & $\underset{\scriptscriptstyle{\pm 1.77}}{19.67}$ & $\underset{\scriptscriptstyle{\pm 2.45}}{31.02}$ & $\underset{\scriptscriptstyle{\pm 1.50}}{20.59}$ & $\underset{\scriptscriptstyle{\pm 1.86}}{32.54}$ & $\underset{\scriptscriptstyle{\pm 1.70}}{17.11}$ & $\underset{\scriptscriptstyle{\pm 4.04}}{56.67}$ & $\underset{\scriptscriptstyle{\pm 1.87}}{22.50}$ \\
     \cmidrule(r){1-1}
     GCN+SSL & $\underset{\scriptscriptstyle{\pm 3.34}}{55.72}$ & $\underset{\scriptscriptstyle{\pm 2.20}}{31.55}$ & $\underset{\scriptscriptstyle{\pm 3.13}}{53.70}$ & $\underset{\scriptscriptstyle{\pm 1.41}}{31.20}$ & $\underset{\scriptscriptstyle{\pm 4.69}}{56.36}$ & $\underset{\scriptscriptstyle{\pm 2.27}}{28.76}$ & $\underset{\scriptscriptstyle{\pm 4.19}}{74.35}$ & $\underset{\scriptscriptstyle{\pm 1.17}}{38.70}$ \\
     \cmidrule(r){1-1}
     GCN+SSL+DropEdge & $\underset{\scriptscriptstyle{\pm 3.99}}{56.56}$ & $\underset{\scriptscriptstyle{\pm 2.75}}{33.19}$ & $\underset{\scriptscriptstyle{\pm 3.50}}{55.87}$ & $\underset{\scriptscriptstyle{\pm 2.96}}{33.57}$ & $\underset{\scriptscriptstyle{\pm 3.17}}{59.57}$ & $\underset{\scriptscriptstyle{\pm 2.68}}{29.17}$ & $\underset{\scriptscriptstyle{\pm 3.88}}{77.64}$ & $\underset{\scriptscriptstyle{\pm 1.43}}{40.89}$ \\
     \cmidrule(r){1-1}
     HyperGCN & $\underset{\scriptscriptstyle{\pm 0.79}}{8.63}$ & $\underset{\scriptscriptstyle{\pm 0.72}}{2.51}$ & $\underset{\scriptscriptstyle{\pm 1.09}}{14.38}$ & $\underset{\scriptscriptstyle{\pm 0.59}}{2.99}$ & $\underset{\scriptscriptstyle{\pm 1.21}}{15.09}$ & $\underset{\scriptscriptstyle{\pm 1.50}}{2.66}$ & $\underset{\scriptscriptstyle{\pm 5.38}}{45.52}$ & $\underset{\scriptscriptstyle{\pm 5.83}}{7.67}$ \\
     \cmidrule(r){1-1}
     HyperGCN+DropEdge & $\underset{\scriptscriptstyle{\pm 1.97}}{18.01}$ & $\underset{\scriptscriptstyle{\pm 0.82}}{13.16}$ & $\underset{\scriptscriptstyle{\pm 1.86}}{31.12}$ & $\underset{\scriptscriptstyle{\pm 0.76}}{15.91}$ & $\underset{\scriptscriptstyle{\pm 2.68}}{32.12}$ & $\underset{\scriptscriptstyle{\pm 2.16}}{19.31}$ & $\underset{\scriptscriptstyle{\pm 4.39}}{61.65}$ & $\underset{\scriptscriptstyle{\pm 5.18}}{23.94}$ \\
     \cmidrule(r){1-1}
     HyperGCN+SSL & $\underset{\scriptscriptstyle{\pm 0.99}}{63.20}$ & $\underset{\scriptscriptstyle{\pm 0.93}}{41.11}$ & $\underset{\scriptscriptstyle{\pm 0.52}}{68.05}$ & $\underset{\scriptscriptstyle{\pm 1.84}}{47.85}$ & $\underset{\scriptscriptstyle{\pm 0.58}}{71.59}$ & $\underset{\scriptscriptstyle{\pm 0.78}}{50.78}$ & $\underset{\scriptscriptstyle{\pm 0.92}}{79.14}$ & $\underset{\scriptscriptstyle{\pm 1.40}}{55.68}$ \\
     \cmidrule(r){1-1}
     HyperGCN+SSL+DropEdge & $\underset{\scriptscriptstyle{\pm 1.50}}{61.35}$ & $\underset{\scriptscriptstyle{\pm 1.70}}{41.62}$ & $\underset{\scriptscriptstyle{\pm 0.19}}{67.31}$ & $\underset{\scriptscriptstyle{\pm 1.64}}{48.42}$ & $\underset{\scriptscriptstyle{\pm 0.46}}{69.59}$ & $\underset{\scriptscriptstyle{\pm 0.70}}{49.31}$ & $\underset{\scriptscriptstyle{\pm 1.72}}{91.82}$ & $\underset{\scriptscriptstyle{\pm 0.09}}{66.29}$ \\
     \cmidrule(r){1-1} 
     {\bf DCAH} & $\underset{\scriptscriptstyle{\pm 1.38}}{16.07}$ & $\underset{\scriptscriptstyle{\pm 0.65}}{7.54}$ & $\underset{\scriptscriptstyle{\pm 2.72}}{15.03}$ & $\underset{\scriptscriptstyle{\pm 0.47}}{4.20}$ & $\underset{\scriptscriptstyle{\pm 3.99}}{19.56}$ & $\underset{\scriptscriptstyle{\pm 0.67}}{4.22}$ & $\underset{\scriptscriptstyle{\pm 3.82}}{39.54}$ & $\underset{\scriptscriptstyle{\pm 1.99}}{6.09}$ \\
     \cmidrule(r){1-1}
     {\bf DCAH}+DropEdge & $\underset{\scriptscriptstyle{\pm 2.41}}{43.66}$ & $\underset{\scriptscriptstyle{\pm 2.35}}{24.76}$ & $\underset{\scriptscriptstyle{\pm 3.82}}{40.95}$ & $\underset{\scriptscriptstyle{\pm 2.94}}{27.24}$ & $\underset{\scriptscriptstyle{\pm 3.50}}{39.87}$ & $\underset{\scriptscriptstyle{\pm 2.91}}{25.88}$ & $\underset{\scriptscriptstyle{\pm 3.46}}{70.34}$ & $\underset{\scriptscriptstyle{\pm 1.78}}{27.22}$ \\
     \cmidrule(r){1-1}
     {\bf DCAH}+SSL & $\underset{\scriptscriptstyle{\pm 0.82}}{69.53}$ & $\underset{\scriptscriptstyle{\pm 1.79}}{46.50}$ & $\underset{\scriptscriptstyle{\pm 0.78}}{70.80}$ & $\underset{\scriptscriptstyle{\pm 0.60}}{50.78}$ & $\underset{\scriptscriptstyle{\pm 0.49}}{82.48}$ & $\underset{\scriptscriptstyle{\pm 1.71}}{57.46}$ & $\underset{\scriptscriptstyle{\pm 3.45}}{88.26}$ & $\underset{\scriptscriptstyle{\pm 0.33}}{66.00}$ \\
     \cmidrule(r){1-1}
     {\bf DCAH}+SSL+DropEdge & $\underset{\scriptscriptstyle{\pm 1.22}}{\bf 72.28}$ & $\underset{\scriptscriptstyle{\pm 1.27}}{\bf 48.82}$ & $\underset{\scriptscriptstyle{\pm 0.37}}{\bf 73.10}$ & $\underset{\scriptscriptstyle{\pm 0.64}}{\bf 52.98}$ & $\underset{\scriptscriptstyle{\pm 0.68}}{\bf 84.78}$ & $\underset{\scriptscriptstyle{\pm 1.08}}{\bf 59.80}$ & $\underset{\scriptscriptstyle{\pm 3.90}}{\bf 93.87}$ & $\underset{\scriptscriptstyle{\pm 0.93}}{\bf 69.79}$ \\

    \bottomrule[1.5pt]
    \end{tabular}}
    \vspace{-0.3cm}
    \label{tab4}
    \end{table}

\subsection{Overall Performance Validation (RQ1)}
\label{4.2}
As the experimental results shown in Table~\ref{tab2}, \ref{tab3} and \ref{tab4}, our DCAH consistently outperforms the GCN and HyperGCN across different E-Commerce datasets in terms of all evaluation metrics, whether enhanced by self-supervised graph pre-training and/or DropEdge or not. This observation validates the superiority of our DCAH method, which can be attributed to: 1) By jointly considering the auxiliary hypergraph, DCAH can not only model the query-item relationship, but also preserve the latent high-order item-item similarity relations. 2) Benefiting from the way that the DCAH combine the bipartite graph and the auxiliary hypergraph. Note that for the HyperGCN method, the original bipartite query-item graph is also utilized since the downstream task is still query-item link prediction. The key difference is that message passing is not performed in the original bipartite graph compared with DCAH. And in the single-channel HyperGCN scenario, the query embeddings are query node features obtained using Byte-Pair Encodings \cite{heinzerling2018bpemb}. And HyperGCN is only adapted to the auxiliary hypergraph to generate the item embeddings. Although the HyperGCN can only generate the item embeddings, they work well on the long-tail part. This could be due to the fact that the auxiliary hypergraph can provide richer information, which is beneficial for the scarce query-item interactions of the long-tail part. For example, some unpopular items from the original bipartite graph may face the cold-start problem, where the GCN method failed to generate robust embeddings, while these items can be updated during HyperGCN training thanks to the high-order item-item relations of the auxiliary hypergraph. In contrast, the performance of the HyperGCN drops drastically on the head part, which also agrees with our intuition that the hypergraph can also introduce some inevitable noise, \textit{e.g.}, a customer’s interest may change while browsing, which can potentially link unrelated items to completely different categories. 

In summary, although the auxiliary hypergraph can provide valuable information, it also introduces the inevitable noise, and our model DCAH, a simple yet effective framework, can jointly capture and adaptively fuse the two complementary information.

\begin{table*}[t]
    \centering
    \caption{Graph smoothness degrees (measured by MAD) with the encoded query/item embeddings by comparing with the different models.}
    \vspace{-0.4cm}
    \resizebox{1\textwidth}{!}{
    \begin{tabular}{c|c|c|c|c|c|c|c|c|c|c|c|c|c}
    \toprule[1.5pt]
    \multirow{1}{*}{\bf Datasets} & {\bf Type} & {\bf GCN} & {\bf GCN+DropEdge} & {\bf GCN+SSL} & {\bf GCN+SSL+DropEdge} & {\bf HyperGCN} & {\bf HyperGCN+DropEdge} & {\bf HyperGCN+SSL} & {\bf HyperGCN+SSL+DropEdge} & {\bf DCAH} & {\bf DCAH+DropEdge} & {\bf DCAH+SSL} & {\bf DCAH+SSL+DropEdge} \\ 
    \midrule[1pt]
    \multirow{2}{*}{E-Commerce 1} & Query & 0.5911 & 0.6234 & 0.6751 & 0.7298 & 0.5436 & 0.5436 & 0.5436 & 0.5436 & 0.6147 & 0.6385 & 0.7776 & 0.8237 \\
    \cmidrule(r){2-14}
    & Item & 0.6438 & 0.6739 & 0.7158 & 0.7439 & 0.6699 & 0.6931 & 0.7396 & 0.7839 & 0.6913 & 0.7246 & 0.7618 & 0.8270 \\
    \midrule[1pt]
    \multirow{2}{*}{E-Commerce 2} & Query & 0.5737 & 0.6022 & 0.6493 & 0.7029 & 0.5216 & 0.5216 & 0.5216 & 0.5216 & 0.5955 & 0.6147 & 0.7567 & 0.8069 \\
    \cmidrule(r){2-14}
    & Item & 0.6273 & 0.6582 & 0.6913 & 0.7254 & 0.6468 & 0.6735 & 0.7197 & 0.7636 & 0.6731 & 0.7064 & 0.7411 & 0.8085 \\
    \midrule[1pt]
    \multirow{2}{*}{E-Commerce 3} & Query & 0.5552 & 0.5813 & 0.6220 & 0.6872 & 0.5023 & 0.5023 & 0.5023 & 0.5023 & 0.5726 & 0.5938 & 0.7488 & 0.7995\\
    \cmidrule(r){2-14}
    & Item & 0.5999 & 0.6327 & 0.6753 & 0.7083 & 0.6273 & 0.6493 & 0.6983 & 0.7458 & 0.6563 & 0.6883 & 0.7208 & 0.7894\\
    
    \bottomrule[1.5pt]
    \end{tabular}}
    \vspace{-0.3cm}
    \label{tab5}
    \end{table*}

\subsection{In Depth Analysis of Hypergraph in Alleviating Data Sparsity (RQ2)}
\label{4.3}
In this section, we easily investigate how the hypergraph helps alleviate the data sparsity. Due to the unique setting of our problem, we partition the items into four different groups by jointly considering their node degrees in bipartite graph and hypergraph. In general, the \textit{head} and \textit{tail1} parts refer to the head part of the original bipartite graph, while the \textit{tail2} and \textit{isolation} parts refer to the long-tail part of the original bipartite graph. From the results shown in Table~\ref{tab2}, \ref{tab3} and \ref{tab4}, we observe that the HyperGCN and DCAH outperform the GCN on the \textit{tail2} and \textit{isolation} parts. Although the HyperGCN only takes the auxiliary hypergraph to generate item embeddings, the superior performance on the long-tail part still shows the potential of the hypergraph in addressing the data sparsity issue. However, due to the inevitable noise of the hypergraph, the performance of the HyperGCN on the \textit{head} and \textit{tail1} part degrades compared to the GCN. Furthermore, in E-Commerce 1 dataset, we find that after adapting the DropEdge strategy, the performance of the HyperGCN on the \textit{head} and \textit{tail1} parts is boosted, while the performance on the \textit{isolation} part drops a little in terms of \textit{MRR}. The possible reason is that the DropEdge may randomly drop some noisy item-item hyperedges, which is beneficial for the \textit{head} and \textit{tail1} parts. While for the \textit{tail2} and \textit{isolation} parts, randomly dropping edges may discard important information of the item-item hyperedges, and these two parts do not contain too much valuable information of the original bipartite graph neither, making the performance worse.

In summary, based on the results, we could conclude that the auxiliary item-item hypergraph has the ability to alleviate the data sparsity issue of the original bipartite graph. Furthermore, if the auxiliary item-item hypergraph is more reliable, \textit{e.g.}, contains less inevitable noise, the benefits will be more. And improving the robustness of the auxiliary item-item hypergraph is our future research direction.

\subsection{Effect of SSL and DropEdge in Addressing Over-Smoothing (RQ3)}
From the Table~\ref{stat}, we find that the three E-Commerce datasets are neither assortative nor disassortative, which indicates that some parts of the original query-item interaction structure are disassortative while the others are assortative. Based on our discussion in Sec~\ref{sec:3.2.2}, we observe the phenomenon that some popular queries/items may make links with unpopular items/queries. Thus, we make an assumption that the head part of the original bipartite graph is more likely to be disassortative while the long-tail part is more likely to be assortative. Then the over-smoothing issue is more likely to be observed in the head part rather than the long-tail part. And from the results of Table~\ref{tab2}, \ref{tab3} and \ref{tab4}, we notice that the performance of the \textit{head} and \textit{tail1} parts are worse than the \textit{tail2} and \textit{isolation} parts, which validates our assumption.

With the self-supervised graph pre-training and DropEdge, the graph-based over-smoothing effect triggered by the disassortative mixing can be alleviated in our framework. To validate the effectiveness of the two strategies in alleviating the over-smoothing effect, in addition to the superior performance produced by our DCAH, we calculate the Mean Average Distance (MAD) \cite{chen2020measuring} over all node embedding pairs learned by the trained DCAH and other three variants: 1) +DropEdge (with the DropEdge); 2) +SSL (with the self-supervised pre-training); 3) +SSL+DropEdge (with the self-supervised pre-training and DropEdge). For complete comparison, we also calculate the MAD for the other methods, GCN and HyperGCN, with their variants. The quantitative MAD metric measures the smoothness of a graph in terms of its node embeddings. The range of MAD is $[0, 1]$, the more closer to 0 the MAD is, the more similar the node representations are to each other, meaning that all the node representations become indistinguishable. The measurement results are shown in Table~\ref{tab5}, where \textit{Query} and \textit{Item} refer to the average similarity score between query nodes and item nodes, respectively. And please note that since the HyperGCN method can only generate the item embeddings, hence, the embedding distance scores of the query nodes are the same.

From the results, we can observe a general trend for the three frameworks: GCN, HyperGCN and DCAH. Take DCAH for example, the DropEdge and SSL can both improve the embedding distance scores, and SSL has better ability at addressing the over-smoothing issue. We guess that is because we use the node-level contrastive loss function in our SSL pre-training, which forces the model to push all the other nodes away from the anchor node, making all the node representations distinguishable.

\section{Conclusions}
In this work, we focus on the query-item prediction problem at E-Commerce stores. Specifically, we discuss the two challenges caused by the original bipartite query-item graph: \textit{long-tail distribution} and \textit{disassortative mixing}. To address these two challenges, we construct an auxiliary item-item hypergraph to augment the original bipartite query-item graph, via using the freely available information from anonymized customer engagement sessions. In the auxiliary hypergraph, the hyperedges contain items present in the same customer sessions. All items engaged by a customer in a single customer session are considered a hyperedge. To tackle the unique setting of the problem, we propose a dual-channel attention-based hypergraph neural network, DCAH, to learn the representations of queries and items by jointly considering the two complementary graphs, the original bipartite graph and the auxiliary hypergraph. Furthermore, we integrate DCAH with self-supervised graph pre-training and the DropEdge training to alleviate disassortative mixing. We evaluate our approach on three randomly sampled anonymized E-Commerce query-item datasets. The experiment results validate the superiority of DCAH. In the future, we may explore more robust methods to construct the hypergraph to reduce the inevitable noise.

% \begin{acks}
% \end{acks}

\newpage
\bibliographystyle{ACM-Reference-Format}

\begin{thebibliography}{10}

\bibitem{gcn}
Thomas~N Kipf and Max Welling.
\newblock Semi-supervised classification with graph convolutional networks.
\newblock {\em arXiv preprint arXiv:1609.02907}, 2016.

\bibitem{graphsage}
William~L Hamilton, Rex Ying, and Jure Leskovec.
\newblock Inductive representation learning on large graphs.
\newblock {\em arXiv preprint arXiv:1706.02216}, 2017.

\bibitem{zhang2018link}
Muhan Zhang and Yixin Chen.
\newblock Link prediction based on graph neural networks.
\newblock {\em Advances in Neural Information Processing Systems},
  31:5165--5175, 2018.

\bibitem{leroy2010cold}
Vincent Leroy, B~Barla Cambazoglu, and Francesco Bonchi.
\newblock Cold start link prediction.
\newblock In {\em Proceedings of the 16th ACM SIGKDD international conference
  on Knowledge discovery and data mining}, pages 393--402, 2010.

\bibitem{ding2021zero}
Hao Ding, Yifei Ma, Anoop Deoras, Yuyang Wang, and Hao Wang.
\newblock Zero-shot recommender systems.
\newblock {\em arXiv preprint arXiv:2105.08318}, 2021.

\bibitem{adhikari2018mining}
Bijaya Adhikari, Parikshit Sondhi, Wenke Zhang, Mohit Sharma, and B~Aditya
  Prakash.
\newblock Mining e-commerce query relations using customer interaction
  networks.
\newblock In {\em Proceedings of the 2018 World Wide Web Conference}, pages
  1805--1814, 2018.

\bibitem{li2018deeper}
Qimai Li, Zhichao Han, and Xiao-Ming Wu.
\newblock Deeper insights into graph convolutional networks for semi-supervised
  learning.
\newblock In {\em Thirty-Second AAAI Conference on Artificial Intelligence},
  2018.

\bibitem{rong2020dropedge}
Yu~Rong, Wenbing Huang, Tingyang Xu, and Junzhou Huang.
\newblock Dropedge: Towards deep graph convolutional networks on node
  classification.
\newblock In {\em International Conference on Learning Representations.
  https://openreview. net/forum}, 2020.

\bibitem{chen2020big}
Ting Chen, Simon Kornblith, Kevin Swersky, Mohammad Norouzi, and Geoffrey~E
  Hinton.
\newblock Big self-supervised models are strong semi-supervised learners.
\newblock {\em Advances in neural information processing systems},
  33:22243--22255, 2020.

\bibitem{cai2020note}
Chen Cai and Yusu Wang.
\newblock A note on over-smoothing for graph neural networks.
\newblock {\em arXiv preprint arXiv:2006.13318}, 2020.

\bibitem{huang2020tackling}
Wenbing Huang, Yu~Rong, Tingyang Xu, Fuchun Sun, and Junzhou Huang.
\newblock Tackling over-smoothing for general graph convolutional networks.
\newblock {\em arXiv preprint arXiv:2008.09864}, 2020.

\bibitem{xia2021self}
Xin Xia, Hongzhi Yin, Junliang Yu, Qinyong Wang, Lizhen Cui, and Xiangliang
  Zhang.
\newblock Self-supervised hypergraph convolutional networks for session-based
  recommendation.
\newblock In {\em Proceedings of the AAAI Conference on Artificial
  Intelligence}, volume~35, pages 4503--4511, 2021.

\bibitem{zimdars2013using}
Andrew Zimdars, David~Maxwell Chickering, and Christopher Meek.
\newblock Using temporal data for making recommendations.
\newblock {\em arXiv preprint arXiv:1301.2320}, 2013.

\bibitem{tuan20173d}
Trinh~Xuan Tuan and Tu~Minh Phuong.
\newblock 3d convolutional networks for session-based recommendation with
  content features.
\newblock In {\em Proceedings of the eleventh ACM conference on recommender
  systems}, pages 138--146, 2017.

\bibitem{wang2020beyond}
Wen Wang, Wei Zhang, Shukai Liu, Qi~Liu, Bo~Zhang, Leyu Lin, and Hongyuan Zha.
\newblock Beyond clicks: Modeling multi-relational item graph for session-based
  target behavior prediction.
\newblock In {\em Proceedings of The Web Conference 2020}, pages 3056--3062,
  2020.

\bibitem{wu2019session}
Shu Wu, Yuyuan Tang, Yanqiao Zhu, Liang Wang, Xing Xie, and Tieniu Tan.
\newblock Session-based recommendation with graph neural networks.
\newblock In {\em Proceedings of the AAAI conference on artificial
  intelligence}, volume~33, pages 346--353, 2019.

\bibitem{wang2021session}
Jianling Wang, Kaize Ding, Ziwei Zhu, and James Caverlee.
\newblock Session-based recommendation with hypergraph attention networks.
\newblock In {\em Proceedings of the 2021 SIAM International Conference on Data
  Mining (SDM)}, pages 82--90. SIAM, 2021.

\bibitem{kherad2020recommendation}
Mahdi Kherad and Amir~Jalaly Bidgoly.
\newblock Recommendation system using a deep learning and graph analysis
  approach.
\newblock {\em arXiv preprint arXiv:2004.08100}, 2020.

\bibitem{shaikh2017recommendation}
Shakila Shaikh, Sheetal Rathi, and Prachi Janrao.
\newblock Recommendation system in e-commerce websites: A graph based
  approached.
\newblock In {\em 2017 IEEE 7th International Advance Computing Conference
  (IACC)}, pages 931--934. IEEE, 2017.

\bibitem{berg2017graph}
Rianne van~den Berg, Thomas~N Kipf, and Max Welling.
\newblock Graph convolutional matrix completion.
\newblock {\em arXiv preprint arXiv:1706.02263}, 2017.

\bibitem{wang2019neural}
Xiang Wang, Xiangnan He, Meng Wang, Fuli Feng, and Tat-Seng Chua.
\newblock Neural graph collaborative filtering.
\newblock In {\em Proceedings of the 42nd international ACM SIGIR conference on
  Research and development in Information Retrieval}, pages 165--174, 2019.

\bibitem{fan2019graph}
Wenqi Fan, Yao Ma, Qing Li, Yuan He, Eric Zhao, Jiliang Tang, and Dawei Yin.
\newblock Graph neural networks for social recommendation.
\newblock In {\em The world wide web conference}, pages 417--426, 2019.

\bibitem{feng2019hypergraph}
Yifan Feng, Haoxuan You, Zizhao Zhang, Rongrong Ji, and Yue Gao.
\newblock Hypergraph neural networks.
\newblock In {\em Proceedings of the AAAI Conference on Artificial
  Intelligence}, volume~33, pages 3558--3565, 2019.

\bibitem{gao2020hypergraph}
Yue Gao, Zizhao Zhang, Haojie Lin, Xibin Zhao, Shaoyi Du, and Changqing Zou.
\newblock Hypergraph learning: Methods and practices.
\newblock {\em IEEE Transactions on Pattern Analysis and Machine Intelligence},
  2020.

\bibitem{wang2020next}
Jianling Wang, Kaize Ding, Liangjie Hong, Huan Liu, and James Caverlee.
\newblock Next-item recommendation with sequential hypergraphs.
\newblock In {\em Proceedings of the 43rd international ACM SIGIR conference on
  research and development in information retrieval}, pages 1101--1110, 2020.

\bibitem{ji2020dual}
Shuyi Ji, Yifan Feng, Rongrong Ji, Xibin Zhao, Wanwan Tang, and Yue Gao.
\newblock Dual channel hypergraph collaborative filtering.
\newblock In {\em Proceedings of the 26th ACM SIGKDD International Conference
  on Knowledge Discovery \& Data Mining}, pages 2020--2029, 2020.

\bibitem{yu2021self}
Junliang Yu, Hongzhi Yin, Jundong Li, Qinyong Wang, Nguyen Quoc~Viet Hung, and
  Xiangliang Zhang.
\newblock Self-supervised multi-channel hypergraph convolutional network for
  social recommendation.
\newblock In {\em Proceedings of the Web Conference 2021}, pages 413--424,
  2021.

\bibitem{van2018representation}
Aaron Van~den Oord, Yazhe Li, and Oriol Vinyals.
\newblock Representation learning with contrastive predictive coding.
\newblock {\em arXiv e-prints}, pages arXiv--1807, 2018.

\bibitem{peng2021self}
Jizong Peng, Ping Wang, Christian Desrosiers, and Marco Pedersoli.
\newblock Self-paced contrastive learning for semi-supervised medical image
  segmentation with meta-labels.
\newblock {\em Advances in Neural Information Processing Systems}, 34, 2021.

\bibitem{cao2021grammatical}
Hannan Cao, Wenmian Yang, and Hwee~Tou Ng.
\newblock Grammatical error correction with contrastive learning in low error
  density domains.
\newblock In {\em Findings of the Association for Computational Linguistics:
  EMNLP 2021}, pages 4867--4874, 2021.

\bibitem{qiu2020gcc}
Jiezhong Qiu, Qibin Chen, Yuxiao Dong, Jing Zhang, Hongxia Yang, Ming Ding,
  Kuansan Wang, and Jie Tang.
\newblock Gcc: Graph contrastive coding for graph neural network pre-training.
\newblock In {\em Proceedings of the 26th ACM SIGKDD International Conference
  on Knowledge Discovery \& Data Mining}, pages 1150--1160, 2020.

\bibitem{you2020graph}
Yuning You, Tianlong Chen, Yongduo Sui, Ting Chen, Zhangyang Wang, and Yang
  Shen.
\newblock Graph contrastive learning with augmentations.
\newblock {\em Advances in Neural Information Processing Systems},
  33:5812--5823, 2020.

\bibitem{xia2022hypergraph}
Lianghao Xia, Chao Huang, Yong Xu, Jiashu Zhao, Dawei Yin, and Jimmy~Xiangji
  Huang.
\newblock Hypergraph contrastive collaborative filtering.
\newblock {\em arXiv preprint arXiv:2204.12200}, 2022.

\bibitem{hu2009disassortative}
Hai-Bo Hu and Xiao-Fan Wang.
\newblock Disassortative mixing in online social networks.
\newblock {\em EPL (Europhysics Letters)}, 86(1):18003, 2009.

\bibitem{broder2000graph}
Andrei Broder, Ravi Kumar, Farzin Maghoul, Prabhakar Raghavan, Sridhar
  Rajagopalan, Raymie Stata, Andrew Tomkins, and Janet Wiener.
\newblock Graph structure in the web.
\newblock {\em Computer networks}, 33(1-6):309--320, 2000.

\bibitem{zhao2021heterogeneous}
Jianan Zhao, Xiao Wang, Chuan Shi, Binbin Hu, Guojie Song, and Yanfang Ye.
\newblock Heterogeneous graph structure learning for graph neural networks.
\newblock In {\em 35th AAAI Conference on Artificial Intelligence (AAAI)},
  2021.

\bibitem{wang2020gcn}
Xiao Wang, Meiqi Zhu, Deyu Bo, Peng Cui, Chuan Shi, and Jian Pei.
\newblock Am-gcn: Adaptive multi-channel graph convolutional networks.
\newblock In {\em Proceedings of the 26th ACM SIGKDD International conference
  on knowledge discovery \& data mining}, pages 1243--1253, 2020.

\bibitem{tang2015line}
Jian Tang, Meng Qu, Mingzhe Wang, Ming Zhang, Jun Yan, and Qiaozhu Mei.
\newblock Line: Large-scale information network embedding.
\newblock In {\em Proceedings of the 24th international conference on world
  wide web}, pages 1067--1077, 2015.

\bibitem{koren2009matrix}
Yehuda Koren, Robert Bell, and Chris Volinsky.
\newblock Matrix factorization techniques for recommender systems.
\newblock {\em Computer}, 42(8):30--37, 2009.

\bibitem{rendle2010factorizing}
Steffen Rendle, Christoph Freudenthaler, and Lars Schmidt-Thieme.
\newblock Factorizing personalized markov chains for next-basket
  recommendation.
\newblock In {\em Proceedings of the 19th international conference on World
  wide web}, pages 811--820, 2010.

\bibitem{liang2016factorization}
Dawen Liang, Jaan Altosaar, Laurent Charlin, and David~M Blei.
\newblock Factorization meets the item embedding: Regularizing matrix
  factorization with item co-occurrence.
\newblock In {\em Proceedings of the 10th ACM conference on recommender
  systems}, pages 59--66, 2016.

\bibitem{eirinaki2005web}
Magdalini Eirinaki, Michalis Vazirgiannis, and Dimitris Kapogiannis.
\newblock Web path recommendations based on page ranking and markov models.
\newblock In {\em Proceedings of the 7th annual ACM international workshop on
  Web information and data management}, pages 2--9, 2005.

\bibitem{wang2022exploiting}
Nan Wang, Shoujin Wang, Yan Wang, Quan~Z Sheng, and Mehmet~A Orgun.
\newblock Exploiting intra-and inter-session dependencies for session-based
  recommendations.
\newblock {\em World Wide Web}, 25(1):425--443, 2022.

\bibitem{wu2021dual}
Longcan Wu, Daling Wang, Kaisong Song, Shi Feng, Yifei Zhang, and Ge~Yu.
\newblock Dual-view hypergraph neural networks for attributed graph learning.
\newblock {\em Knowledge-Based Systems}, 227:107185, 2021.

\bibitem{bretto2013hypergraph}
Alain Bretto.
\newblock Hypergraph theory.
\newblock {\em An introduction. Mathematical Engineering. Cham: Springer},
  2013.

\bibitem{newman2003mixing}
Mark~EJ Newman.
\newblock Mixing patterns in networks.
\newblock {\em Physical review E}, 67(2):026126, 2003.

\bibitem{yan2021two}
Yujun Yan, Milad Hashemi, Kevin Swersky, Yaoqing Yang, and Danai Koutra.
\newblock Two sides of the same coin: Heterophily and oversmoothing in graph
  convolutional neural networks.
\newblock {\em arXiv preprint arXiv:2102.06462}, 2021.

\bibitem{heinzerling2018bpemb}
Benjamin Heinzerling and Michael Strube.
\newblock {BPEmb: Tokenization-free Pre-trained Subword Embeddings in 275
  Languages}.
\newblock In {\em Proceedings of the Eleventh International Conference on
  Language Resources and Evaluation (LREC 2018)}, Miyazaki, Japan, May 7-12,
  2018 2018. European Language Resources Association (ELRA).

\bibitem{chen2020simple}
Ming Chen, Zhewei Wei, Zengfeng Huang, Bolin Ding, and Yaliang Li.
\newblock Simple and deep graph convolutional networks.
\newblock {\em arXiv preprint arXiv:2007.02133}, 2020.

\bibitem{hu2020open}
Weihua Hu, Matthias Fey, Marinka Zitnik, Yuxiao Dong, Hongyu Ren, Bowen Liu,
  Michele Catasta, and Jure Leskovec.
\newblock Open graph benchmark: Datasets for machine learning on graphs.
\newblock {\em Advances in neural information processing systems},
  33:22118--22133, 2020.

\bibitem{chen2020measuring}
Deli Chen, Yankai Lin, Wei Li, Peng Li, Jie Zhou, and Xu~Sun.
\newblock Measuring and relieving the over-smoothing problem for graph neural
  networks from the topological view.
\newblock In {\em Proceedings of the AAAI Conference on Artificial
  Intelligence}, volume~34, pages 3438--3445, 2020.

\end{thebibliography}

\appendix
\end{document}